\documentclass[preprint,showpacs,amsmath,amssymb,floats,nofootinbib]{revtex4}
\usepackage{graphicx}
\usepackage[english]{babel}
\usepackage{amsmath}
\usepackage{amssymb}
\usepackage{color}

\newcommand{\vk}{\mathbf{k}}

\newcommand{\be}{\begin{eqnarray}}
\newcommand{\ee}{\end{eqnarray}}

\renewcommand{\vr}{\mathbf{r}}

\begin{document}

\title{Analytic Solution of the \textit{Ornstein-Zernike} Relation for Inhomogeneous Liquids}

\author{Yan He}
\affiliation{College of Physical Science and Technology, Sichuan University, Chengdu, Sichuan 610064, PRC}
\author{Stuart A. Rice}
\email{sarice@uchicago.edu}
\affiliation{The James Franck Institute and Department of Chemistry, The University of Chicago, Chicago, IL 60637, USA}
\author{Xinliang Xu}
\email{xinliang@csrc.ac.cn}
\affiliation{Complex Systems Division, Beijing Computational Science Research Center, Beijing 100193, PRC}

\begin{abstract}
The properties of a classical simple liquid are strongly affected by application of an external potential that supports inhomogeneity. To understand the nature of these property changes the equilibrium particle distribution functions of the liquid have, typically, been calculated directly using either integral equation or density functional based analyses.
In this study we develop a different approach with a focus on two distribution functions that characterize the inhomogeneous liquid: the pair direct correlation function $c(\vr_1,\vr_2)$ and the pair correlation function $g(\vr_1,\vr_2)$. With $g(\vr_1,\vr_2)$ considered to be an experimental observable, we solve the \textit{Ornstein-Zernike} equation for the inhomogeneous liquid to obtain $c(\vr_1,\vr_2)$, using information about the well studied and resolved $g^{(0)}(\vr_1,\vr_2)$ and $c^{(0)}(\vr_1,\vr_2)$ for the parent homogeneous ($^{(0)}$) system. In practical cases, where $g(\vr_1,\vr_2)$ is available from experimental data in discrete form, the resulting $c(\vr_1,\vr_2)$ is expressed as an explicit function of $g(\vr_1,\vr_2)$ in discrete form. A weaker continuous form of solution is also obtained, in the form of an integral equation with finite integration limits. The result obtained with our formulation is tested against the exact solutions for the correlation and distribution functions of a one-dimensional inhomogeneous hard rod liquid. Following the success of that test the formalism is extended to higher dimensional systems with explicit consideration of the two-dimensional liquid.
\end{abstract}

\maketitle

\section{introduction}

When placed under the influence of an external potential, the structure of an otherwise homogeneous fluid can be changed significantly\cite{Nygard,Chandler,Dietrich}, leading to qualitative changes in system properties\cite{Mittal}. From a formal point of view, the external potential can describe a boundary condition, such as contact of the fluid with a wall, or can represent the influence of one particle on those that surround it in a bulk homogeneous fluid.  Consequently, the study of inhomogeneous fluids is of importance to the understanding of diverse phenomena ranging from hydrophobicity\cite{Berne,Stanley}, to protein structures\cite{Levy}, to freezing transitions\cite{Brader,Schmidt}, to glass transitions\cite{Lang}, to the structure and properties of ordinary fluids\cite{Nygard1,Henderson-book}.

The overwhelming majority of theoretical studies of inhomogeneous fluids use integral equations\cite{Caccamo} or density functional theory\cite{Rosenfeld}.  The integral equation approach is based on direct calculation of the pair correlation function of the fluid, $g(\vr_1,\vr_2)$ from equations obtained from one or the other truncation of the BBGKY hierarchy, e.g. by partial summation of the diagrammatic representation in terms of particle interactions.  Alternatively, the \emph{Ornstein-Zernike} (OZ) equation\cite{OZ} can be coupled with closure relations\cite{PY,Rowlinson} that introduce an approximate analytic relationship between the pair correlation function $g(\vr_1,\vr_2)$ and the pair direct correlation function $c(\vr_1,\vr_2)$, or introduce a simple functional form for the so-called bridge function.  While good results can be achieved for the properties of certain homogeneous model fluids, especially with closure relations such as the generalized-mean-spherical approximation\cite{Waisman}, the Rogers-Young\cite{Rogers} approximation, the second order Percus-Yevick approximation\cite{Henderson1,Henderson2}, and modified hypernetted chain approximations\cite{RA}, great care is needed in order to extend this approach to inhomogeneous fluids\cite{Lang,Sarman,Ishizuka} and to high density fluids close to crystallization\cite{Brader}.  The density functional theory approach, particularly in the modification known as fundamental measure theory\cite{Rosenfeld}, approaches the calculation of fluid properties by constructing a functional of the excess free energy through weighted densities\cite{RSLT,Ta1,Ta2}.  By improving the underlying equation of state of the fluid\cite{Santos,Hansen}, the theory proves to be very successful in generating accurate predictions of the properties of both the pure hard sphere fluid and hard-sphere mixtures.  However, fundamental measure theory is designed to account for excluded volume effects in a hard sphere system, and extension to systems with other interactions is very difficult.

The pair correlation function is determined experimentally from measurements of the angular distribution of scattered radiation, and can be considered an observable property of the fluid.  The pair direct correlation function, although defined in terms of the pair correlation function, is an inferred, not an observed, property of the fluid.  For a homogeneous fluid that is translation invariant and isotropic, the OZ equation can be studied in Fourier space where an algebraic relation between the Fourier transforms of the pair direct and pair correlation functions, $\hat{c}(\vk)$ and $\hat{h}(\vk)$, can be obtained, namely $\hat{c}(\vk)=\hat{h}(\vk)/[1+\rho\hat{h}(\vk)]$ .  But for an inhomogeneous fluid finding $c(\vr_1,\vr_2)$ as an explicit function of $g(\vr_1,\vr_2)$ is not a trivial problem.

In this paper we show how the pair direct correlation function of an inhomogeneous fluid can be calculated in terms of the pair correlation function without the use of closure approximations of the type described above.  The calculation follows the point of view that the pair distribution function of the homogeneous fluid is an observable.  Then, assuming that for systems that are influenced by external potentials that are local and short ranged, far from the external potential $g(\vr_1,\vr_2)$ reduces to its homogeneous value $g^{(0)}(\vr_1-\vr_2)$, we solve the OZ equation for $c(\vr_1,\vr_2)$ using the observed $g(\vr_1,\vr_2)$ as input, and thereby obtain $c(\vr_1,\vr_2)$ as a function of $g(\vr_1,\vr_2)$.  It is worth noting that finding this functional relationship provides the exact amount of information needed for some problems.  For example, in many theoretical treatments within the density functional theory framework the crystallization transition is described by the instability of the liquid with respect to a specified crystal structure that is characterized by a set of reciprocal lattice vectors\cite{Xu1,Xu2}.  As a result, the behavior of the set of direct correlation functions, defined as  functional derivatives of the system excess free energy with respect to the density distribution at crystallization, can be predicted at those points in Fourier space identified by the reciprocal lattice vectors.  In many cases these predicted values cannot be directly verified or applied since the direct correlation functions are not directly measurable, but it is sufficient for indirect verification to translate the predictions to values of the observable pair correlation function.

The rest of this paper is organized in the following fashion: in Sec.\ref{sec-1D} we show the reduction of $g(\vr_1,\vr_2)$ to its homogeneous value $g^{(0)}(\vr_1,\vr_2)$ far from the location of the external potential for a specific one dimensional model system. Then using the observed $g(\vr_1,\vr_2)$ as input, which is typically available from experiment in discrete form, we analytically solve the OZ equation and obtain $c(\vr_1,\vr_2)$ as an explicit discrete function of $g(\vr_1,\vr_2)$. In Sec.\ref{sec-1Dc} we show that with an additional weak assumption concerning the asymptotic behavior of the continuous pair direct correlation function this discrete solution transforms into a weaker but more straightforward solution obtained from an integral equation with finite integration limits, which allows numerical evaluation at an arbitrary level of precision when an analytical representation of $g(\vr_1,\vr_2)$ is provided as input. In Sec.\ref{sec-num} we compare our results with results obtained from the known exact solution for a one-dimensional inhomogeneous hard rod liquid. The good agreement achieved even by our numerical results from the weaker solution obtained in Sec.\ref{sec-1Dc}, shows that for this particular model system the approximations we have made are very reasonable. In Sec.\ref{sec-hD} the formalism is generalized to apply to systems in greater than one-dimension. In Sec.\ref{sec-end} we discuss the relation between the solution in discrete form and the solution in continuous form and the conditions where they apply.

\section{Solving the \textit{Ornstein-Zernike} equation in one dimension}
\label{sec-1D}

In its most general form, the OZ equation serves as an implicit relation between the pair direct correlation function and the pair correlation function given by
\be
h(\vr_1,\vr_2)=c(\vr_1,\vr_2)+\int d\vr_3\rho(\vr_3)c(\vr_1,\vr_3)h(\vr_3,\vr_2)
\ee
where $h(\vr_1,\vr_2)=g(\vr_1,\vr_2)-1$ and $\rho(\vr_3)$ is the number density at $\vr_3$.  For a homogeneous fluid that is translation invariant and isotropic, the OZ equation can be studied in Fourier space where an algebraic relation between $\hat{c}(\vk)$ and $\hat{h}(\vk)$ can be obtained as $\hat{c}(\vk)=\hat{h}(\vk)/[1+\rho\hat{h}(\vk)]$ . But for an inhomogeneous fluid (e.g. liquid under the influence of an external potential), finding $c(\vr_1,\vr_2)$ is not a trivial problem.

We proceed as follows. To simplify the OZ relation we introduce the scaled correlation functions $H(\vr_1,\vr_2)=\sqrt{\rho(\vr_1)\rho(\vr_2)}h(\vr_1,\vr_2)$ and $C(\vr_1,\vr_2)=\sqrt{\rho(\vr_1)\rho(\vr_2)}c(\vr_1,\vr_2)$. Then the OZ equation transforms into the following integral equation:
\be
H(\vr_1,\vr_2)=C(\vr_1,\vr_2)+\int d\vr_3 H(\vr_1,\vr_3)C(\vr_3,\vr_2) \label{HC}
\ee
Our goal is to obtain $C(\vr_1,\vr_2)$ from a known function of $H(\vr_1,\vr_2)$. Due to the external field, there is no translation invariance, and $H(\vr_1,r_2)$ does not depend only on $\vr_1-\vr_2$. The integral equation Eq.(\ref{HC}) is a standard Fredholm equation of the second kind. For one-dimensional systems, this integral equation can be discretized into matrix form
\be
H_{ij}=C_{ij}+\Delta x\sum_{k=-\infty}^{\infty}H_{ik}C_{kj},\quad (-\infty<i,j<\infty)\label{dHC}
\ee
Here $H_{ij}=H(x_i,x_j)$, $C_{ij}=C(x_i,x_j)$ and $x_i=i\Delta x$.
Eq.(\ref{dHC}) can also be rewritten in a more convenient form as follows:
\be
\sum_{k=-\infty}^{\infty}(\delta_{ik}+\Delta x H_{ik})(\delta_{kj}-\Delta x C_{kj})=\delta_{ij}\label{mHC}
\ee
For convenience, we introduce matrices $\mathcal{A}$ and $\mathcal{B}$ with matrix elements $\mathcal{A}_{ij}=\delta_{ij}+\Delta x H_{ij}$ and $\mathcal{B}_{ij}=\delta_{ij}-\Delta x C_{ij}$, the difficulty associated with the solution of Eq.(\ref{mHC}) can be seen as we rewrite it in a 3 by 3 block form:

\be
\mathcal{A}=
\left(\begin{array}{ccc}
A_{11} & A_{12} & A_{13}\\
A_{21} & A_{22} & A_{23}\\
A_{31} & A_{32} & A_{33}
\end{array}
\right),\quad
\mathcal{B}=
\left(\begin{array}{ccc}
B_{11} & B_{12} & B_{13}\\
B_{21} & B_{22} & B_{23}\\
B_{31} & B_{32} & B_{33}
\end{array}
\right)\label{subAB}
\ee
where $A_{ab}$ and $B_{ab}$ with $a,b=1,2,3$ are sub-matrices of $\mathcal{A}$ and $\mathcal{B}$. The matrix element $\mathcal{A}_{ij}$ ($\mathcal{B}_{ij}$) belongs to sub-matrix $A_{ab}$ ($B_{ab}$) under the following conditions: $a=1$ or $a=2$ or $a=3$ for $i<-N$ or $-N<i<N$ or $i>N$, respectively, and $b=1$ or $b=2$ or $b=3$ for $j<-N$ or $-N<j<N$ or $j>N$, respectively, with $N$ determined by $N\Delta x=R$.

\begin{figure}
\centerline{\includegraphics[clip,width=0.3\textwidth]{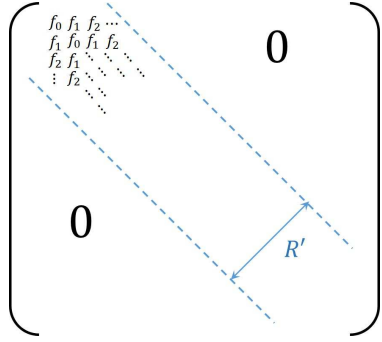}}
\caption{Elements in matrix $\mathcal{A}$ for a homogeneous system. Matrix $\mathcal{A}$ is a Toeplitz matrix where matrix elements are constant along the diagonal direction with $\mathcal{A}_{ij}=f(|i-j|)$. Here the length scale $R'$ indicates the range of particle pair correlation as a result of short-ranged their interactions.}\label{fig1}
\end{figure}

Taking the inverse of the above block matrix, Eq.(\ref{mHC}) leads to
\be
&&B_{22}=(A_{22}-\delta A_{22})^{-1}\neq A_{22}^{-1}\label{BA}\\
&&\delta A_{22}=A_{21}\cdot \tilde{A}^{-1}_{31}\cdot \tilde{A}_{32}
+A_{23}\cdot \tilde{A}^{-1}_{13}\cdot \tilde{A}_{12},\quad
\tilde{A}_{ab}=A^{-1}_{1a}A_{1b}-A^{-1}_{3a}A_{3b}\nonumber
\ee
Without a straightforward way to carry out the inversion of $(A_{22}-\delta A_{22})$, it is natural for us to try to generate successive approximations to $B_{22}$ as the inverse of $A_{22}$ with increasing range $R$, expecting that the effect of $\delta A_{22}$ will eventually decay away at some finite value of $R$. However, for reasons that will be clear in the later discussion, it can be shown that there exists a finite length scale $R$ on which the elements in all blocks of matrix $\mathcal{A}$ except $A_{22}$ reduce to the values of the matrix elements for the corresponding homogeneous system, which can be represented by a Toeplitz matrix\cite{Kadanoff} (Figure \ref{fig1}). As illustrated, for a uniform system these matrix elements are constants along the diagonal with $\mathcal{A}_{ij}=f(|i-j|)$, where $f(|x|)$ decays to 0 at a characteristic length scale $R'$. As a result, there exists no straightforward way to find a cutoff length where the effect of $\delta A_{22}$ decays to zero. In other words, there will always be a finite level of error associated with $B_{22}$ when evaluated with information from matrix $\mathcal{A}$ within a finite space, namely $A^{-1}_{22}$. For homogeneous systems it is well known that we can bypass this difficulty by utilizing the fact that $\mathcal{A}_{ij}$ is constant along the diagonal direction, hence can be evaluated with a Fourier transformation. For inhomogeneous systems that method is no longer applicable as $\mathcal{A}_{ij}$ is no longer constant along the diagonal direction. A new method has to be developed for accurate evaluation of the integral over the whole space required in Eq.(\ref{HC}), or the evaluation of $B_{22}$ in Eq.(\ref{BA}).

For many problems of interest, the external potential is local. For these systems we now develop a systematic way to obtain the analytic solution of Eq.(\ref{HC}) by reduction of the range of integration to a finite region.

For simplicity, we consider a one-dimensional classical fluid of $M+1$ identical particles with pair additive potential $V_{int}(x_i,x_j)=V_{int}(|x_i-x_j|)$ between particles centered at $x_i$ and $x_j$, respectively. The generalization of this treatment to systems in higher dimensions will be presented in section \ref{sec-hD}.

We write $\rho_n^0(x_1,\cdots,x_n)$ for the $n$-particle probability functions of the homogeneous system in the absence of the external potential. By fixing one of these identical particles at the origin (labeled as particle 0), we can study the $n$-particle probability functions $\rho_n^{\phi}(x_1,\cdots,x_n)$ for the rest of the particles, which now constitute an inhomogeneous system in the local force field due to particle 0. By relating the $M$ particle term of the grand partition function for the inhomogeneous case to the $M+1$ particle term for the homogeneous case\cite{SL}, it can be shown that $\rho^{\phi}_1(x_1)=\rho^{(0)}_2(0,x_1)/\rho^{(0)}_1(0)$ and $\rho^{\phi}_2(x_1,x_2)=\rho_3^{(0)}(0,x_1,x_2)/\rho^{(0)}_1(0)$. From the definition of the pair distribution function $g(x_1,x_2)=\rho_2(x_1,x_2)/[\rho_1(x_1)\rho_1(x_2)]$, we have
\be
g^{\phi}(x_1,x_2)=\frac{t^{(0)}(0,x_1,x_2)}{g^{(0)}(|x_1|)g^{(0)}(|x_2|)}\label{g-phi}
\ee
given that $\rho_1^{(0)}=\rho_0$ and $g^{(0)}(0,x)=g^{(0)}(|x|)$ for the homogeneous system, and
$t^{(0)}(0,x_1,x_2)=\frac{\rho_3^{(0)}(0,x_1,x_2)}{\rho^{(0)}_1(0)\rho^{(0)}_1(x_1)\rho^{(0)}_1(x_2)}$ the triplet distribution function.

For a short ranged potential $V_{int}(x)$, we expect a finite characteristic length scale $R'$ beyond which the effect of the existence of a particle at the origin is negligible. Then for $x_1\gg R'$, $x_2$ cannot be simultaneously close enough to feel both the particle at 0 and the particle at $x_1$. With the help of Eq. (7), we have $h(x_1,x_2)=0$ when $x_2$ is far from $x_1$, $\rho_1(x_1)=\rho_1(x_2)=\rho_0$ and $h(x_1,x_2)=h^{(0)}(|x_1-x_2|)$ when $x_2$ is far from 0. That is, by setting $R\gg R'$ we expect $H(x_1,x_2)=\sqrt{\rho(x_1)\rho(x_2)}h(x_1,x_2)$ to reduce to $H_0(x_1-x_2)=\rho_0h^{(0)}(|x_1-x_2|)$ for either $|x_1|>R$ or $|x_2|>R$.

Utilizing this property, we can solve Eq.(\ref{HC}) for the inhomogeneous case by comparison with the corresponding functions for the homogeneous case. For homogeneous systems we have $H(x_1,x_2)=H_0(x_1-x_2)$ and $C(x_1,x_2)=C_0(x_1-x_2)$, and Eq.(\ref{HC}) in matrix form becomes
\be
\sum_{k=-\infty}^{\infty}[\delta_{ik}+\Delta x (H_0)_{ik}][\delta_{kj}-\Delta x (C_0)_{kj}]=\delta_{ij}
\ee
As before, we introduce $\mathcal{A}^{(0)}_{ik}=\delta_{ik}+\Delta x (H_0)_{ik}$ and $\mathcal{B}^{(0)}_{ik}=\delta_{ik}-\Delta x (C_0)_{ik}$, which have the following block form
\be
\mathcal{A}^{(0)}=
\left(\begin{array}{ccc}
A_{11}^{(0)} & A_{12}^{(0)} & A_{13}^{(0)}\\
A_{21}^{(0)} & A_{22}^{(0)} & A_{23}^{(0)}\\
A_{31}^{(0)} & A_{32}^{(0)} & A_{33}^{(0)}
\end{array}
\right),\quad
\mathcal{B}^{(0)}=
\left(\begin{array}{ccc}
B_{11}^{(0)} & B_{12}^{(0)} & B_{13}^{(0)}\\
B_{21}^{(0)} & B_{22}^{(0)} & B_{23}^{(0)}\\
B_{31}^{(0)} & B_{32}^{(0)} & B_{33}^{(0)}
\end{array}
\right)\label{subAB0}
\ee
leading to the result in the same form:
\be
(B^{(0)}_{22})^{-1}=A^{(0)}_{22}-A^{(0)}_{21}\cdot (\tilde{A}^{(0)}_{31})^{-1}\cdot \tilde{A}^{(0)}_{32}
-A^{(0)}_{23}\cdot (\tilde{A}^{(0)}_{13})^{-1}\cdot \tilde{A}^{(0)}_{12}
\label{BA0}
\ee

As discussed, for an inhomogeneous system by setting $R\gg R'$ we expect $H(x_1,x_2)$ to reduce to $H_0(x_1-x_2)$ for $|x_1|>R$ or $|x_2|>R$, leading to $\mathcal{A}=\mathcal{A}^{(0)}$ for all blocks except that $A_{22}\neq A^{(0)}_{22}$. Then by taking the difference between Eqs.(\ref{BA}) and (\ref{BA0}) we find
\be
B_{22}^{-1}-(B_{22}^{(0)})^{-1}=A_{22}-A_{22}^{(0)}\label{AB1}
\ee
wherein all quantities involved are confined to the finite space defined by $|x_1|<R$ and $|x_2|<R$. Using the definitions of sub-matrices $A_{22}$, $A^{(0)}_{22}$, $B_{22}$ and $B^{(0)}_{22}$ in Eq.(\ref{subAB}) and Eq.(\ref{subAB0}), we can move the $(B^{(0)}_{22})^{-1}$ term in Eq.(\ref{AB1}) to the right hand side and obtain $c(x_1, x_2)$ as an explicit discrete function of $g(x_1, x_2)$, $g^{(0)}(x_1, x_2)$ and $c^{(0)}(x_1, x_2)$.

\section{ONE DIMENSIONAL SOLUTION IN CONTINUOUS FORM}
\label{sec-1Dc}

Before considering the verification of the approximation leading to Eq.(\ref{AB1}) by comparison with exact results for a model system, we examine the physical meaning of Eq.(\ref{AB1}) by converting it back to an integral equation. Multiplying both sides of Eq.(\ref{AB1}) by $B^{(0)}_{22}$ on the right, with help of Eq.(\ref{BA}) we have
\be
(A_{22}-A^{(0)}_{22})\cdot B^{(0)}_{22}=(A_{22}-\delta A_{22})\cdot(B^{(0)}_{22}-B_{22})\label{AB1p}
\ee
Since $\mathcal{A}=\mathcal{A}^{(0)}$ for all blocks except $A_{22}$, the only non-zero elements of the $2N\times 2N$ matrix of $\delta A_{22}$ appear in the $n\times n$ block located at the top left and bottom right corners characterized by $N-n<|i|,|j|<N$, where $n$ is defined through $R'=n\delta x$. We now make the additional approximation that, like the behavior of $H(x_1,x_2)$, the unknown $C(x_1,x_2)$ we are trying to determine also reduces to its homogeneous counterpart $C_0(x_1-x_2)=\rho_0 c^{(0)}(x_1-x_2)$ for $|x_1|>R-R'$ or $|x_2|>R-R'$. If so, then we have $(\mathcal{B}^{(0)}-\mathcal{B})_{ij}=0$ for $N-n<|i|<N$ or $N-n<|j|<N$, leading to $\delta A_{22}(B^{(0)}_{22}-B_{22})=0$. Under this assumption we see that Eq. (\ref{AB1p}) is greatly reduced to
\be
(A_{22}-A^{(0)}_{22})\cdot B^{(0)}_{22}=A_{22}\cdot(B^{(0)}_{22}-B_{22})
\label{AB2}
\ee
Making use of the definition of matrices $\mathcal{A}$ and $\mathcal{B}$ displayed in Eq.(\ref{subAB}), Eq.(\ref{AB2}) becomes
\be
\sum_{j=-N}^N\Big(H_{ij}-(H_0)_{ij}\Big)\Big(\delta_{jk}-\Delta x (C_0)_{jk}\Big)
=\sum_{j=-N}^N\Big(\delta_{ij}+\Delta x H_{ij}\Big)\Big(C_{jk}-(C_0)_{jk}\Big)\nonumber
\ee
which in the continuous limit is just
\be
&&H(x_1,x_2)-C(x_1,x_2)-\int_{-R}^R dx_3 H(x_1,x_3)C(x_3,x_2)\nonumber\\
&&\qquad=H_0(x_1-x_2)-C_0(x_1-x_2)-\int_{-R}^R dx_3 H_0(x_1-x_3)C_0(x_3-x_2)
\label{HC1}
\ee
where $H_0$ and $C_0$ are the known scaled pair correlation and direct correlation functions for the homogeneous system. Eq. (\ref{HC1}) shows that, under the conditions $H(x_1,x_2)=H_0(x_1-x_2)$ and $C(x_1,x_2)=C_0(x_1-x_2)$ for either $|x_1|>R$ or $|x_2|>R$, the OZ relation has been reduced to a finite space, at the price of introducing an extra homogeneous term that characterizes the finite size effect of the integral term $\int dx_3H(x_1,x_3)C(x_3,x_2)$.

It is worthwhile noting that, Eq.(\ref{HC1}) appears naturally in case $\int_R^{\infty}dx_3H(x_1,x_3)C(x_3,x_2)=\int_R^{\infty}dx_3H_0(x_1-x_3)C_0(x_3-x_2)$, which follows as a direct result from the conditions $H(x_1,x_2)=H_0(x_1-x_2)$ and $C(x_1,x_2)=C_0(x_1-x_2)$ for either $|x_1|>R$ or $|x_2|>R$. That is, these conditions are sufficient for the derivation of Eq.(\ref{HC1}), and our lengthy derivation from Eq.(\ref{dHC}) to Eq.(\ref{AB2}) is just to show that these conditions are also necessary.

\section{Test and verification of the approximation}
\label{sec-num}

To test our scheme for determining the pair direct correlation function in terms of the pair correlation function in an inhomogeneous liquid we consider a one-dimensional classical fluid of hard rods of length $a$. Exact solutions for the pair direct correlation function and the pair correlation function of the homogeneous one-dimensional hard rod liquid and the one-dimensional hard rod liquid under the influence of a spatially varying external potential $\phi$ are known\cite{Kirkwood,Lieb}. We consider the simple inhomogeneous liquid generated when one hard rod is fixed at the origin. For convenience, we first list the well-known analytic results.

\subsection{The homogeneous 1D hard rod liquid}

For a homogeneous suspension of one dimensional hard rods at number density $\rho_0$, the pair correlation function $g^{(0)}(x_1-x_2)$ and the pair direct correlation function $c^{(0)}(x_1-x_2)$ have the forms\cite{Cui,Zernike}
\be
&&g^{(0)}(x_1-x_2)=\sum_{k=1}^{\infty}\frac{\eta^{k-1}}{(1-\eta)^k}\frac{(y-k)^{k-1}}{(k-1)!}
e^{-\frac{\eta(y-k)}{1-\eta}}\Theta(y-k)\\
&&c^{(0)}(x_1-x_2)=\frac{-1}{1-\eta}-\frac{\eta(1-y)}{(1-\eta)^2},\qquad y\leq 1\\
&&c^{(0)}(x_1-x_2)=0,\qquad\qquad\qquad\qquad y>1 \nonumber
\ee
respectively, where $y=|x_1-x_2|/a$, $\eta=\rho a$, and $\Theta(x)$ is the step function. Then the scaled functions $H_0(x_1-x_2)$ and $C_0(x_1-x_2)$ as used in Eq.(\ref{HC1}) can be obtained through the definitions $H_0(x_1-x_2)=\rho_0[g^{(0)}(x_1-x_2)-1]$ and $C_0(x_1-x_2)=\rho_0c^{(0)}(x_1-x_2)$.

\subsection{The inhomogeneous 1D hard rod liquid}

For an inhomogeneous suspension of mobile one dimensional hard rods moving under the potential arising from an additional identical hard rod centered and fixed at the origin, analytical results can be obtained for the number density $\rho^{\phi}(x)$ and pair correlation function $g^{\phi}(x_1,x_2)$ by relating the $M$ particle term of the grand partition function for the inhomogeneous liquid to the $M+1$ particle term for the homogeneous liquid\cite{SL}. It is found that $\rho^{\phi}(x)=\rho_0g^{(0)}(0,x)$ and $g^{\phi}(x_1,x_2)$ is given by Eq.(\ref{g-phi}). The one dimensional homogeneous hard rod system is unique in that the triplet correlation function is exactly represented as a product of pair correlation functions for adjacent pairs of particles\cite{Kirkwood}:
\be
t^{(0)}(x_1,x_2,x_3)=g^{(0)}(x_1,x_2)g^{(0)}(x_2,x_3)\qquad\mbox{for}\quad x_1<x_2<x_3
\ee
For the case that interactions are restricted to nearest neighbors, Percus\cite{Percus} has shown that the pair direct correlation function $c^{\phi}(x_1,x_2)$ of a suspension of one dimensional hard rods under the influence of an external potential $\phi$ has the form
\be
&&c^{\phi}(x_1,x_2)=\frac{-1}{1-\int_{x_2-a}^{x_2}\rho^{\phi}(w)dw}
-\int_{x_2}^{x_1+a}\frac{\rho^{\phi}(z)dz}{[1-\int_{z-a}^z\rho^{\phi}(w)dw]^2},\qquad y\leq 1\\
&&c^{\phi}(x_1,x_2)=0,\qquad\qquad\qquad\qquad y>1 \nonumber
\ee
where $y=|x_1-x_2|/a$. Then the scaled functions $H(x_1,x_2)$ and $C(x_1,x_2)$ as used in Eq.(\ref{HC1}) can be obtained through $H(x_1,x_2)=\sqrt{\rho^{\phi}(x_1)\rho^{\phi}(x_2)}(g^{\phi}(x_1,x_2)-1)$ and $C(x_1,x_2)=\sqrt{\rho^{\phi}(x_1)\rho^{\phi}(x_2)}c^{\phi}(x_1,x_2)$.

\subsection{Verification of Eq.(\ref{HC1})}

Given the analytical forms of $H_0(x_1,x_2)$, $C_0(x_1,x_2)$, and $H(x_1,x_2)$ we can numerically calculate $C_{num}(x_1,x_2)$ through Eq.(\ref{HC1}). We shall compare the results of that calculation with the results obtained from the analytical form of $C(x_1,x_2)$. To carry out the numerical calculation, we write Eq.(\ref{HC1}) in discrete form using quantities defined in Eq.(\ref{dHC}):
\be
H_{ij}-C_{ij}-\Delta x\sum_{k=-N}^N H_{ik}C_{kj}
=H^{(0)}_{ij}-C^{(0)}_{ij}-\Delta x\sum_{k=-N}^N H^{(0)}_{ik}C^{(0)}_{kj}\nonumber
\ee
With the input of $2N\times 2N$ matrices $H$, $H^{(0)}$, and $C^{(0)}$, by simple linear algebra the $2N\times 2N$ matrix $C$ can be written in the form
\be
C=(I+\Delta x H)^{-1}\cdot(H-H^{(0)}+C^{(0)}+\Delta x H^{(0)}\cdot C^{(0)})\nonumber
\ee
where $I$ is the $2N\times 2N$ identity matrix.

\begin{figure}
\centerline{\includegraphics[clip,width=0.9\textwidth]{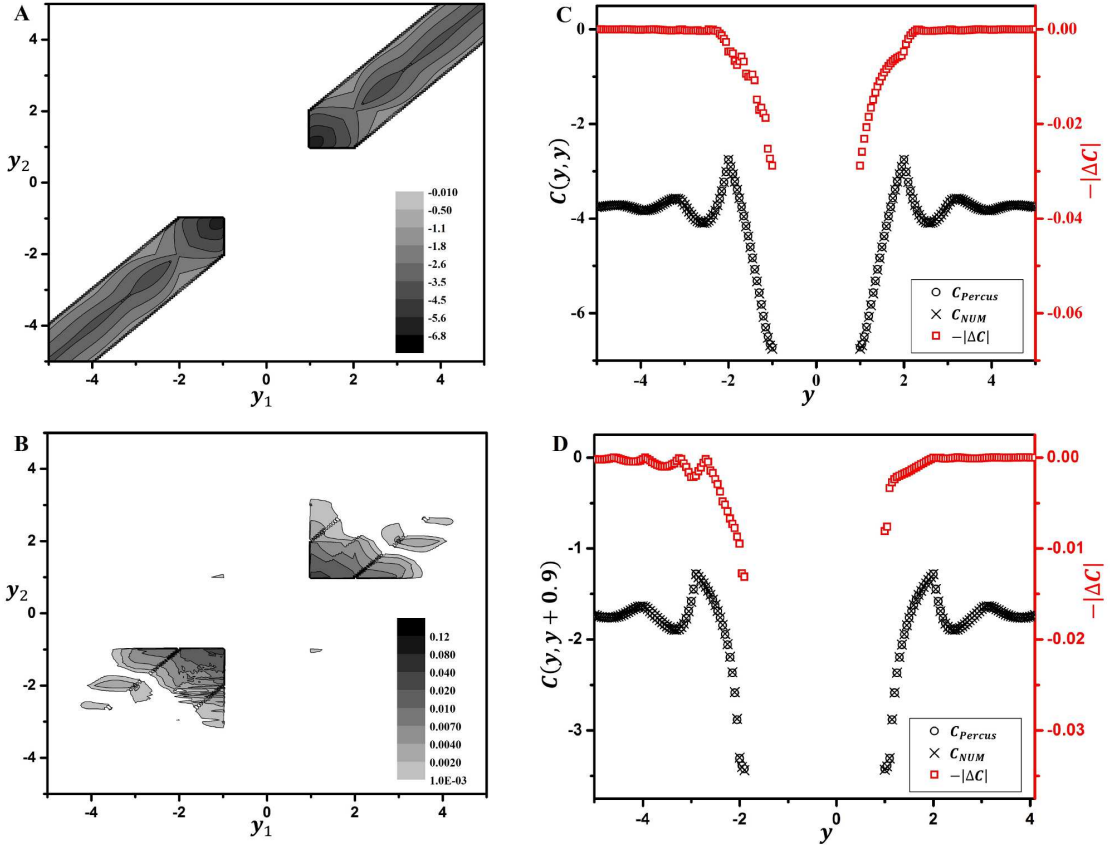}}
\caption{Using scaled length $y=x/a$, we show (A) the analytical $C(y_1,y_2)$ obtained by Percus\cite{Percus}; (B) the amplitude of the difference between our numerical results $C_{num}(y_1,y_2)$ and the analytical results $C(y_1,y_2)$ predicted by Percus, $|\Delta C(y_1,y_2)|=|C_{num}(y_1,y_2)-C(y_1,y_2)|$; (C) the analytical $C(y_1,y_2)$ by Percus, our numerical result $C_{num}(y_1,y_2)$, and $-|\Delta C(y_1,y_2)|$ along the $y_2=y_1$ line; (D) the analytical $C(y_1,y_2)$ by Percus, our numerical result $C_{num}(y_1,y_2)$, and $-|\Delta C(y_1,y_2)|$ along the $y_2=y_1+0.9$ line. Please note that for comparison purposes, the scales for $-|\Delta C(y_1,y_2)|$ in (C) and (D) are designated to be 100 times smaller than the corresponding scales for $C(y_1,y_2)$ and $C_{num}(y_1,y_2)$.}\label{fig2}
\end{figure}

At low packing fractions, the average particle spacing is large and the scaled pair distribution functions are relatively simple, i.e. have little structure. Setting $\Delta x=0.001a$ and $R=10a$, we find that the pair direct correlation function $C_{num}(x_1,x_2)$ can be obtained through Eq.(\ref{HC1}) with high precision even at the high packing fraction $\eta=\rho_0a=0.61$. Indeed, comparison with the exact $C(x_1,x_2)$ obtained by Percus\cite{Percus}, shows that $C_{num}(x_1,x_2)$ obtained through Eq.(\ref{HC1}) is very accurate, as characterized by the small value of $\Delta C(x_1,x_2)=C_{num}(x_1,x_2)-C(x_1,x_2)$ over the range $-5a<x_1,x_2<5a$ (Figure \ref{fig2}A, \ref{fig2}B). The reader should note that the scales used in Figure 2B for $\Delta C(x_1,x_2)$ are much smaller than the scales used in Figure 2A for $C(x_1,x_2)$. To better illustrate the comparison, we show the analytical $C(x_1,x_2)$, our numerical result $C_{num}(x_1,x_2)$, and $-|\Delta C(x_1,x_2)|$ along the $x_1=x_2$ line (Figure \ref{fig2}C), and along the $x_1=x_2+0.9a$ line (Figure \ref{fig2}D). The comparison shows excellent agreement between previous analytical results and our numerical results in both Figure \ref{fig2}C and Figure \ref{fig2}D. The differences, characterized by $-|\Delta C(x_1,x_2)|$ and plotted with scales that are 100 times smaller than the corresponding scales for $C(x_1,x_2)$ and $C_{num}(x_1,x_2)$ for comparison purpose, can be shown to be tiny in most parameter regions, with the biggest relative error much less than $1\%$.

\section{THE \textit{ORNSTEIN-ZERNIKE} equation in higher dimensions}
\label{sec-hD}

\begin{figure}
\centerline{\includegraphics[clip,width=0.3\textwidth]{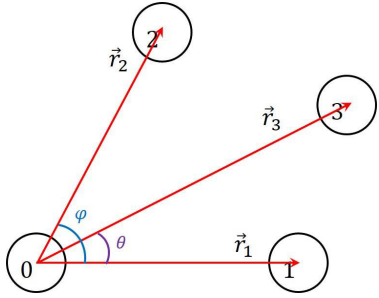}}
  \caption{An inhomogeneous system in the force field due to particle 0. The location of the n-th particle is characterized by $\vr_n$. We denote $\varphi$ as the angle between $\vr_1$ and $\vr_2$, $\theta$ as the angle between $\vr_1$ and $\vr_3$.}\label{fig3}
\end{figure}

In this section, we generalize the formalism of section \ref{sec-1D} to higher dimensional cases. Specifically, we fold the $|r|>R$ effects into a finite space. We will show a 2D example and the generalization procedure to 3D should be similar. In cases where the external potential is local, our strategy is the same as for the one-dimensional case. Consider a two-dimensional classical fluid of $M+1$ identical particles with pair additive potential $V_{int}(\vr_1,\vr_2)=V_{int}(|\vr_1-\vr_2|)$ between particles centered at $\vr_1$ and $\vr_2$, respectively. Again by fixing one of these identical particles at the origin (labeled as particle 0), we can study the properties of the rest of the particles as an inhomogeneous system in the local force field due to particle 0. Following a procedure similar to that used in section \ref{sec-1D}, we have: $H(\vr_1,\vr_2)=H(r_1,r_2,\varphi)$ and $C(\vr_1,\vr_2)=C(r_1,r_2,\varphi)$, where $\varphi$ is the angle between $\vr_1$ and $\vr_2$ (Figure \ref{fig3}). The scaled OZ relation Eq.(\ref{HC}) transforms into:
\be
H(r_1,r_2,\varphi)=C(r_1,r_2,\varphi)+\int_0^{2\pi}d\theta\int_0^{\infty}
r_3dr_3H(r_1,r_3,\theta)C(r_3,r_2,\theta-\varphi)
\label{HC2d}
\ee
where $\theta$ is the angle between $\vr_1$ and $\vr_3$ (figure \ref{fig3}).

Since $H(r_1,r_2,\varphi)=H(r_1,r_2,2\pi+\varphi)=H(r_1,r_2,-\varphi)$ and $C(r_1,r_2,\varphi)=C(r_1,r_2,2\pi+\varphi)=C(r_1,r_2,-\varphi)$. we have
\be
H(r_1,r_2,\varphi)=\sum_0^{\infty}H_m(r_1,r_2)\cos(m\varphi)\quad
C(r_1,r_2,\varphi)=\sum_0^{\infty}C_m(r_1,r_2)\cos(m\varphi)\label{Hm}
\ee
where $H_m(r_1,r_2)=\frac{1}{\pi}\int_0^{2\pi}H(r_1,r_2,\varphi)\cos(m\varphi)d\varphi$ and $C_m(r_1,r_2)=\frac{1}{\pi}\int_0^{2\pi}C(r_1,r_2,\varphi)\cos(m\varphi)d\varphi$.

Rewriting Eq.(\ref{HC2d}) with the help of Eq.(\ref{Hm}), it can be shown that for all $m$ we have:
\be
H_m(r_1,r_2)=C_m(r_1,r_2)+\int_0^{\infty}r_3dr_3H_m(r_1,r_3)C_m(r_3,r_2)
\label{HCm}
\ee
With scaled functions $\tilde{H}_m(r_1,r_2)=\sqrt{r_1r_2}H_m(r_1,r_2)$ and $\tilde{C}_m(r_1,r_2)=\sqrt{r_1r_2}C_m(r_1,r_2)$, Eq.(\ref{HCm}) transforms into the following simple one-dimensional integral equation:
\be
\tilde{H}_m(r_1,r_2)=\tilde{C}_m(r_1,r_2)+\int_0^{\infty}dr_3\tilde{H}_m(r_1,r_3)\tilde{C}_m(r_3,r_2)
\label{HCm1}
\ee
Eq.(\ref{HCm1}) is in the same form as Eq.(\ref{HC}). Following a similar procedure to that used before for a short ranged potential $V_{int}(x)$, we expect a finite characteristic length scale $R'$ beyond which the effect of the existence of a particle at the origin is negligible. Then for $r_1\gg R'$, $\vr_2$ cannot be simultaneously close enough to feel both the particle at 0 and the particle at $\vr_1$. Since $h(r_1,r_2,\varphi)=0$ when $\vr_2$ is far from $\vr_1$, $\rho(\vr_1)=\rho(\vr_2)=\rho_0$ and $h(\vr_1,\vr_2)=h_0(|\vr_1-\vr_2|)$ when $\vr_2$ is far from 0, we have $\tilde{H}_m(r_1,r_2)=\tilde{H}^{(0)}_m(r_1,r_2)$ for either $r_1>R$ or $r_2>R$, where $\tilde{H}^{(0)}_m(r_1,r_2)=\frac{\sqrt{r_1r_2}}{\pi}\int_0^{2\pi}\rho_0h_0(r_1,r_2,\varphi)\cos(m\varphi)d\varphi$ can be obtained from the parent homogeneous system. Going through the same procedure as introduced earlier (from Eq.(\ref{dHC}) to Eq.(\ref{BA0})), results similar to Eq.(\ref{AB1}) can be obtained as:
\be
((B_m)_{22})^{-1}-((B^{(0)}_m)_{22})^{-1}=(A_m)_{22}-(A^{(0)}_m)_{22}\label{ABm}
\ee
where all quantities involved are confined to the finite space defined by $r_1<R$ and $r_2<R$. Note that Eq.(\ref{ABm}) has the same form as Eq.(\ref{AB1}) for the one-dimensional case.

Adopting the argument we used in section \ref{sec-1Dc}, if we make the further assumption that, like the behavior of $H(\vr_1,\vr_2)$, the unknown $C(\vr_1,\vr_2)$ we are trying to determine also reduces to its homogeneous counterpart $C_0(\vr_1,\vr_2)=\rho c^{(0)}(|\vr_1-\vr_2|)$ for either $r_1>R$ or $r_2>R$, we find
\be
\int_{|\vr_3|>R}d\vr_3 H(\vr_1,\vr_3)C(\vr_3,\vr_2)=\int_{|\vr_3|>R}d\vr_3 H_0(\vr_1,\vr_3)C_0(\vr_3,\vr_2)
\ee
which, used together with Eq.(\ref{HC2d}), gives
\be
&&H(\vr_1,\vr_2)-C(\vr_1,\vr_2)-\int_{|\vr_3|<R}d\vr_3 H(\vr_1,\vr_3)C(\vr_3,\vr_2)\nonumber\\
&&\qquad=H_0(\vr_1,\vr_2)-C_0(\vr_1,\vr_2)-\int_{|\vr_3|<R}d\vr_3 H_0(\vr_1,\vr_3)C_0(\vr_3,\vr_2)\label{HC-hD}
\ee
Note that Eq.(\ref{HC-hD}) has the same form as Eq.(\ref{HC1}) for the one-dimensional case.

\section{discussion}
\label{sec-end}

For homogeneous systems Eq.(\ref{HC}) can be readily solved by resorting to Fourier transformation, which utilizes the fact that the system is translation invariant hence the elements in matrices $\mathcal{A}$ and $\mathcal{B}$ are constants along the diagonal. For inhomogeneous systems such translation invariance no longer exists, and finding a solution of Eq.(\ref{HC}) becomes a challenge. We have shown that, under certain conditions, analytical solution of Eq.(\ref{HC}) with the unknown $C(\vr_1,\vr_2)$ as a function of $H(\vr_1,\vr_2)$ within a finite region is possible.

Our derivation illustrates that as long as there exists a finite length scale $R$ on which $H(\vr_1,\vr_2)$ reduces to $H_0(\vr_1,\vr_2)=\rho_0 h^{(0)}(\vr_1-\vr_2)$ for $|\vr_1| > R$ or $|\vr_2| > R$, the solution can be obtained explicitly. That solution is expressed in discretized fashion in Eq.(\ref{AB1}) for one dimensional problems and in Eq.(\ref{ABm}) for higher dimensional problems, regardless of whether the resulting $C(\vr_1,\vr_2)$ is short-ranged or not. Since in many model systems the pair direct correlation function is shorter-ranged than the pair correlation function, we can presume that the unknown $C(\vr_1,\vr_2)$ also reduces to its homogeneous counterpart $C_0(\vr_1,\vr_2)=\rho_0 c^{(0)}(\vr_1-\vr_2)$ for $|\vr_1| > R$ or $|\vr_2| > R$, leading to a solution in the continuous form of an integral equation with finite integration limits, shown in Eq.(\ref{HC1}) for one dimensional problems and Eq.(\ref{HC-hD}) for higher dimensional problems. While solution in this latter form is more straightforward, it is weaker in the sense that the resulting $C(\vr_1,\vr_2)$ has to be double-checked with the ``short-ranged'' presumption, which is not a priori guaranteed to be valid. In cases where it is not valid, e.g. in systems with ``screening'' the pair direct correlation function can be long-ranged while the pair correlation function remains short-ranged\cite{SL}. In this case our derivation in Sec. III shows that the solution in the continuous fashion (Eq.(\ref{HC1})) no longer applies, while the solution in discrete form (Eq.(\ref{AB1})) is still valid. In other words, Eq.(\ref{AB1}) is valid for all model systems with a finite length scale R where $H(\vr_1,\vr_2)$ reduces to $H_0(\vr_1,\vr_2)$ for $|\vr_1| > R$ or $|\vr_2| > R$, while Eq.(\ref{HC1}) is only valid for those model systems with a finite length scale R where, for $|\vr_1| > R$ or $|\vr_2| > R$, both $H(\vr_1,\vr_2)$ and $C(\vr_1,\vr_2)$ need to reduce to $H_0(\vr_1,\vr_2)$ and $C_0(\vr_1,\vr_2)$, respectively.

It should be noted that the discrete form of solution is as accurate as is the continuous form, since the two solutions are equivalent for model systems where both are valid. Ideally, if an analytical representation of $g(\vr_1, \vr_2)$ is provided as input, both the solution in discrete form and the solution in continuous form can be evaluated numerically with arbitrary precision. In practice, the input $g(\vr_1, \vr_2)$ is available from experimental data in discrete form, with finite measurement resolution. Then, the solution in discrete form or in continuous form can be numerically evaluated consistent with that resolution level. And, although we have shown that the proposed approach to calculating the pair direct correlation function works very well for systems with ``hard sphere'' type of interactions, and it is plausible that it will work well for other systems with short ranged interactions, it remains necessary to generalize this treatment to systems with other forms of local external potentials to affirm that plausibility.

\section*{Acknowledgement}

We thank M. Tchernookov for helpful discussions. Y. H. acknowledges the support by National Natural Science Foundation of China \#11404228, S. A. R. acknowledges the support of this research by the NSF MRSEC at the University of Chicago DMR-1420709, X. X. acknowledges the support by National Natural Science Foundation of China \#11575020, and \#U1530401.

\end{document}